\documentclass[twocolumn,showpacs,preprintnumbers,amsmath,amssymb]{revtex4}
\usepackage[dvips]{graphicx}
\usepackage{dcolumn}
\usepackage{bm}    

\begin{document}
\title{About Landau-Hopf scenario in a system of coupled self-oscillators}
\author{Alexander~P.~Kuznetsov, Sergey~P.~Kuznetsov, Igor~R.~Sataev and Ludmila~L.~Turukina}
\affiliation{Kotel'nikov's Institute of Radio-Engineering and Electronics of RAS, Saratov Branch,\\
Zelenaya 38, Saratov, 410019, Russian Federation.}

\date{\today}

\begin{abstract}
The conditions are discussed for which an ensemble of interacting oscillators may demonstrate the Landau-Hopf scenario of successive birth of multi-frequency quasi-periodic motions. A model is proposed that is a network of five globally coupled oscillators characterized by controlled degree of activation of individual oscillators. Illustrations are given for successive birth of tori of increasing dimension via quasi-periodic Hopf bifurcations.
\end{abstract}


\maketitle

\section{Introduction}

In this paper we discuss a problem that attracts attention for many years, but has not received yet a satisfactory solution and understanding. It consists in a question of possibility of observation of the scenario of transition to complex dynamical behavior suggested by Landau and Hopf, e.g. in the context of hydrodynamic turbulence. This scenario includes successive secondary quasi-periodic Hopf bifurcation, which follow the primary one, being accompanied with a sequential birth of incommensurable frequency components of the motion. We suggest here some novel approach to this long-standing problem based on artificial constructing and examination of model systems with the required type of behavior, which allow physical implementation (say, in mechanics or electronics). A concrete example we propose is composed as a network of coupled van der Pol oscillators. It may be thought as a setup universal in some physical sense, due to the fact that the van der Pol oscillator is associated with a normal form for the Andronov-Hopf bifurcation.

Back in the forties of the last century, Landau~\cite{1}, and later Hopf~\cite{2}, suggested that in the course of variation of the control parameter (Reynolds number) the hydrodynamic turbulence may appear via the following sequence of events. First, after the stability loss of a steady flow, a self-oscillatory regime appears that corresponds to a limit cycle in the phase space of the system. (This transition is called now the Andronov-Hopf bifurcation.) Then, in turn, the periodic motion becomes unstable and undergoes the secondary bifurcation. (In modern terminology it is referred to as the Neimark-Sacker bifurcation.) Now, the dynamics takes place on a two-dimensional torus in the phase space of the system, born as a result of the soft bifurcation transition. Next, this motion undergoes a similar bifurcation, resulting in emergence of one more oscillatory component, with its own independent frequency, giving rise to a three-frequency torus, and so on. As a result of successfully growing number of the incommensurable frequencies in the spectrum of oscillations, the regime is complicating, and finally the turbulence develops.

An attractive feature of the Landau argumentation is its generality; in fact it does not appeal to a specific physical nature or to concrete mathematical equations of the system, and, hence, must have equal relation to a variety of multidimensional and spatially extended dissipative nonlinear systems.

Later the approach of Landau-Hopf was criticized in a frame of the concept of Ruelle and Takens~\cite{3}, who indicated a possibility of destruction of low-dimensional tori due to arbitrarily small variations in the system evolution operator accompanied with the onset of a strange attractor corresponding to chaotic dynamics.

To verify the Ruelle-Takens concept, several researchers performed numerical experiments for model systems and observed that the low-dimensional tori usually survive under small perturbations, although these tori may be destroyed giving birth to chaos at sufficiently large magnitude of them~\cite{4,5}. Hence, a widespread conclusion of the universality of the Ruelle-Takens scenario should be regarded as hasty and ill-founded.

On the other hand, one can criticize the Landau-Hopf picture in a frame of the concept of synchronization, which is a universal nonlinear phenomenon~\cite{6}. Indeed, in situations of close basic frequencies for the involved oscillatory components (or of some combinations of their frequencies), the system may demonstrate either attractive periodic orbits, or lower dimension attractive resonant tori placed on higher-dimensional tori~\cite{7,8,9,10,11}. So, resonances and the possible occurrence of chaos are expected to modify essentially the picture of dynamical regimes and bifurcations; it makes the Landau-Hopf scenario problematic. In this context it would be interesting nevertheless to search for specific systems, in which such a scenario actually could occur, at least, one could observe a sufficiently large number of initial quasi-periodic bifurcations. Constructing an appropriate model based on a set of coupled nonlinear oscillators is the purpose of the present article.

When choosing a model to analyze, we take into account several important physical aspects. First, the oscillatory modes responsible for the dynamics must be characterized by different degrees of activation. If we consider, say, an ensemble of coupled van der Pol oscillators, we must introduce a set of parameters $\lambda _{i} $ controlling the negative friction and select them appropriately to ensure conditions for gradual involving of the modes in the motion in the course of decrease of dissipation level. (In the context of the hydrodynamic problems it just corresponds to increasing Reynolds number.) Second, all the oscillatory modes have to be separated in frequency in sufficient degree. Otherwise, essential interaction of the modes would occur, which can destroy the Landau-Hopf picture. Third and finally, it is desirable to have a situation, where the coupled oscillators are arranged in such way that they all are involved essentially in the interaction; it means that no preferable interaction of each concrete partial oscillator should occur, say, with spatially close neighbors, or with different number of the neighbors. In such situation the control parameter of each single oscillator $\lambda _{i} $ will regulate a certain quasi-periodic bifurcation. Otherwise, it might happen that due to the different number of interacting neighbors the elements with close level of activation would have essentially different influence on the picture of the multi-frequency dynamics~\cite{11}.

Selecting the model, it is crucial to guarantee the occurrence of the definite type of the bifurcations. For the Landau-Hopf scenario these must be bifurcations of soft (non-catastrophic) birth of tori of increasing dimension. (In modern literature such bifurcations are usually referred to as the quasi-periodic Hopf bifurcations~\cite{12,13}.) In contrast, resonance mode locking phenomena are associated with saddle-node bifurcations of tori. Unfortunately, no simple numerical algorithms for identifying the quasi-periodic bifurcations are elaborated. Qualitatively, the types of bifurcations of quasi-periodic regimes can be identified in computations by means of analysis of the behavior of the Lyapunov exponents, and this method will be used henceforth.

\section{Network of five coupled self-oscillators}

Let us construct a model satisfying the above requirements. It will be a kind of network of five oscillators with global coupling (that implies coupling of each element with each other). We suppose the oscillators to have equidistant spectrum of natural frequencies and leave a single frequency parameter $\Delta $ in the model, which controls the mutual mismatch between the oscillators. For large values of this parameter the oscillators will be desynchronized. We assume that all oscillators are characterized by different values of the parameters responsible for exceeding the excitation threshold; it will allow gradual activation of the corresponding modes in the course of the transition under consideration.

The set of equations for the model system we propose reads

\begin{equation} \label{eq1_} \begin{array}{l} {\ddot{x}-(\lambda _{1} -x^{2} )\dot{x}+x+\frac{\mu }{4} (4\dot{x}-\dot{y}-\dot{z}-\dot{w}-\dot{v})=0,} \\ {\ddot{y}-(\lambda _{2} -y^{2} )\dot{y}+(1+\frac{\Delta }{4} )y+\frac{\mu }{4} (4\dot{y}-\dot{x}-\dot{z}-\dot{w}-\dot{v})=0,} \\ {\ddot{z}-(\lambda _{3} -z^{2} )\dot{z}+(1+\frac{\Delta }{2} )z+\frac{\mu }{4} (4\dot{z}-\dot{y}-\dot{x}-\dot{w}-\dot{v})=0,} \\ {\ddot{w}-(\lambda _{4} -w^{2} )\dot{w}+(1+\frac{3\Delta }{4} )w+\frac{\mu }{4} (4\dot{w}-\dot{y}-\dot{x}-\dot{z}-\dot{v})=0,} \\ {\dot{v}-(\lambda _{5} -v^{2} )\dot{v}+(1+\Delta )v+\frac{\mu }{4} (4\dot{v}-\dot{y}-\dot{x}-\dot{z}-\dot{w})=0.} \end{array} \end{equation}

Here $\lambda_{i}$ are control parameters responsible for excitation of the partial oscillators, $\Delta $ determines the frequency detuning of the oscillators, and the frequency of the first oscillator is unit. We set hereafter$\, \, \lambda _{1} =0.1,\, \, \lambda _{2} =0.2,\, \, \lambda _{3} =0.3$, $\, \lambda _{4} =0.4,\, \, \lambda _{5} =0.5$.

Our main illustration will be a Lyapunov chart on the parameter plane for frequency detuning $\Delta $ and dissipative coupling $\mu $. To draw this chart we proceed as follows~\cite{10,11}. First, select a point on the parameter plane and compute the spectrum of Lyapunov exponents $\Lambda _{i} $ for the Poincar\'{e} map of the system (\ref{eq1_}) there. Then, analyzing the largest four exponents, take into account their sings and evaluate a number of zero exponents among them to determine a type of the dynamical regime (the attractor type) in the system. Namely, it is determined as

\begin{enumerate}
\item  a limit cycle \textit{P}, if$\, \, \Lambda _{1} <0,\; \Lambda {}_{2} <0,\; \Lambda _{3} <0,\, \Lambda _{4} <0$,

\item  a two-frequency torus $T_{2} $, if $\Lambda _{1} =0,\; \Lambda {}_{2} <0,\; \Lambda _{3} <0,\, \Lambda _{4} <0$,

\item  a three-frequency torus $T_{3} $, if $\Lambda _{1} =0,\; \Lambda {}_{2} =0,\; \Lambda _{3} <0{\rm ,}\, \Lambda _{4} <0$,

\item  a four-frequency torus $T_{4} $, if $\Lambda _{1} =0,\; \Lambda {}_{2} =0,\; \Lambda _{3} =0,\, \Lambda _{4} <0$,

\item  a five-frequency torus $T_{5} $, if $\Lambda _{1} =0,\; \Lambda {}_{2} =0,\; \Lambda _{3} =0,\, \Lambda _{4} =0$,

\item  chaos \textit{C}, if$\, \, \Lambda _{1} >0,\; \Lambda {}_{2} <0,\; \Lambda _{3} <0,\, \Lambda _{4} <0$.
\end{enumerate}

Then, the respective pixel on the parameter plane is attributed with a definite color, in accordance with the detected dynamical regime. Scanning the entire parameter plane, we get the two-parameter chart of the dynamical regimes.

\begin{figure}
\includegraphics[width=0.45\textwidth,keepaspectratio]{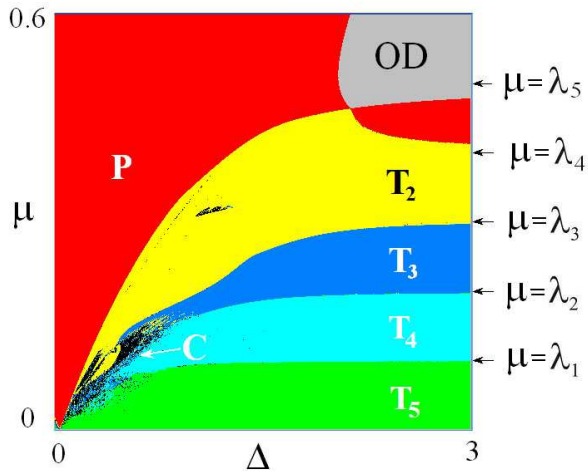}
\caption{Chart of Lyapunov exponents for the network of five globally coupled identical van der Pol oscillators (\ref{eq1_}), where $\lambda _{1} =0.1,\, \, \lambda _{2} =0.2,\, \, \lambda _{3} =0.3,\, \, \, \lambda _{4} =0.4,\, \, \lambda _{5} =0.5.$}
\label{f1}
\end{figure}

Figure~\ref{f1} shows the Lyapunov chart obtained for the model (\ref{eq1_}) on the parameter plane$\, \, (\Delta ,\mu )$. Because of the structure of the system, the main resonance effects are actually excluded. In particular, no tongues of resonant tori of different dimensions similar to those mentioned e.g. in Refs.~\cite{10,11} are observed. The only pronounced tongue corresponding to a resonant two-frequency torus may be seen in the region$\, \, \Delta \le 0.5$, between the domains of the five-frequency tori and of the complete synchronization. Interestingly, it is immersed in the region of chaos that occurs at small coupling \footnote{ Occurrence of chaos for weak coupling in small networks was noted e.g. in Ref.~\cite{14}, and there this effect was named the "phase chaos".}. Here the three-frequency, four-frequency, and partially the five-frequency tori are destroyed, although the chaos is weak. (Typical values of the positive Lyapunov exponent are of order $10^{-3}-10^{-2}$ here.) To some extent, one can say that the Ruelle-Takens scenario occurs in this region. Note additionally that a small island of chaos takes place surrounded by the area of two-frequency tori, at the coupling coefficient values about 0.3-0.4. Hence, to give rise to chaos it is not needed to deal necessarily with destruction of three-frequency tori; it appears to be sufficient to have the smaller dimension.

On the other hand, in the case of large detuning of the oscillators,$\, \, \Delta \ge 1$, with decreasing dissipation parameter $\mu $ one can observe the appearance of all the tori of higher dimension. Boundaries of the relevant areas in the asymptotic $\Delta \to \infty $ correspond to the values of the control parameters $\mu $=$\lambda_{i}$. (For convenience, these values are indicated by arrows on the right edge of the chart.) Thus, a decrease in the parameter of dissipative coupling qualitatively corresponds to the pattern expected for the Landau-Hopf scenario.

\section{The cascade of quasi-periodic bifurcations}

\begin{figure}
\includegraphics[width=0.45\textwidth,keepaspectratio]{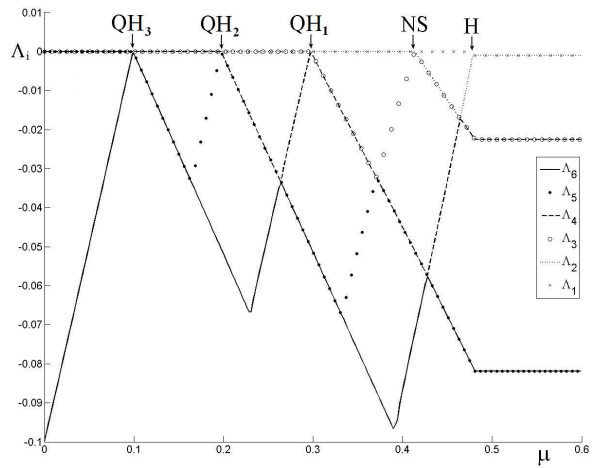}
\caption{Graphs of six Lyapunov exponents of the Poincar\'{e} map of model (\ref{eq1_}) at the value of the detuning parameter $\Delta $=3.}
\label{f2}
\end{figure}

To be confident that we deal here really with a cascade of quasi-periodic Hopf-Neimark-Sacker bifurcations for tori of higher dimensions, we turn to the graphs of the Lyapunov exponents for the Poincar\'{e} map of the system in Fig.~\ref{f2}. The exponents are plotted versus the coupling parameter $\mu $, along the vertical line $\Delta =3$ in the parameter plane. One can see that for large values of the coupling parameter, all the exponents are negative that corresponds to the so-called oscillator death regime (OD). This mode is typical for coupled oscillators, and occurs due to the dissipative nature of coupling: at sufficiently strong couplings it dampens the oscillations~\cite{6}. The point \textit{H} corresponds to the Andronov-Hopf bifurcation, and here one exponent $\Lambda _{1} $ becomes equal zero. Then, at the Neimark-Sacker bifurcation \textit{NS}, one more exponent $\Lambda _{2} $ vanishes. At the point $QH_{1} $ the next exponent $\Lambda _{3} $ becomes zero that corresponds to the birth of the three-dimensional torus. To identify the nature of the bifurcation, we note that before the bifurcation the exponents $\Lambda _{3} $ and $\Lambda _{4} $ are equal. This is a characteristic feature of the soft quasi-periodic Neimark-Sacker bifurcation~\cite{12}. Alternatively, at the quasi-periodic saddle-node bifurcation the Lyapunov exponents behave differently~\cite{12,13}.

Then, exactly in the same manner the remaining exponents behave: at the point $QH_{2} $ a four-dimensional torus appears due to the secondary quasi-periodic Neimark-Sacker bifurcation, and at the point $QH_{3} $ a five-dimensional torus arises.

\begin{figure}
\includegraphics[width=0.45\textwidth,keepaspectratio]{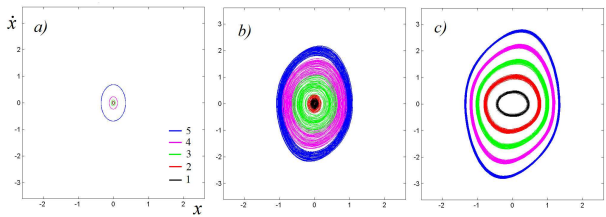}
\caption{Evolution of the phase portraits of oscillators of the system (\ref{eq1_}) with decrease of the dissipation parameter: a) $\mu =0.45$, the limit cycle, b) $\mu =0.25$, the three-frequency torus, c)~$\mu =0.05$, the five-frequency torus. The numbers numerate partial oscillators of the system; the frequency detuning parameter is $\Delta =3$.}
\label{f3}
\end{figure}

\begin{figure}
\includegraphics[width=0.45\textwidth,keepaspectratio]{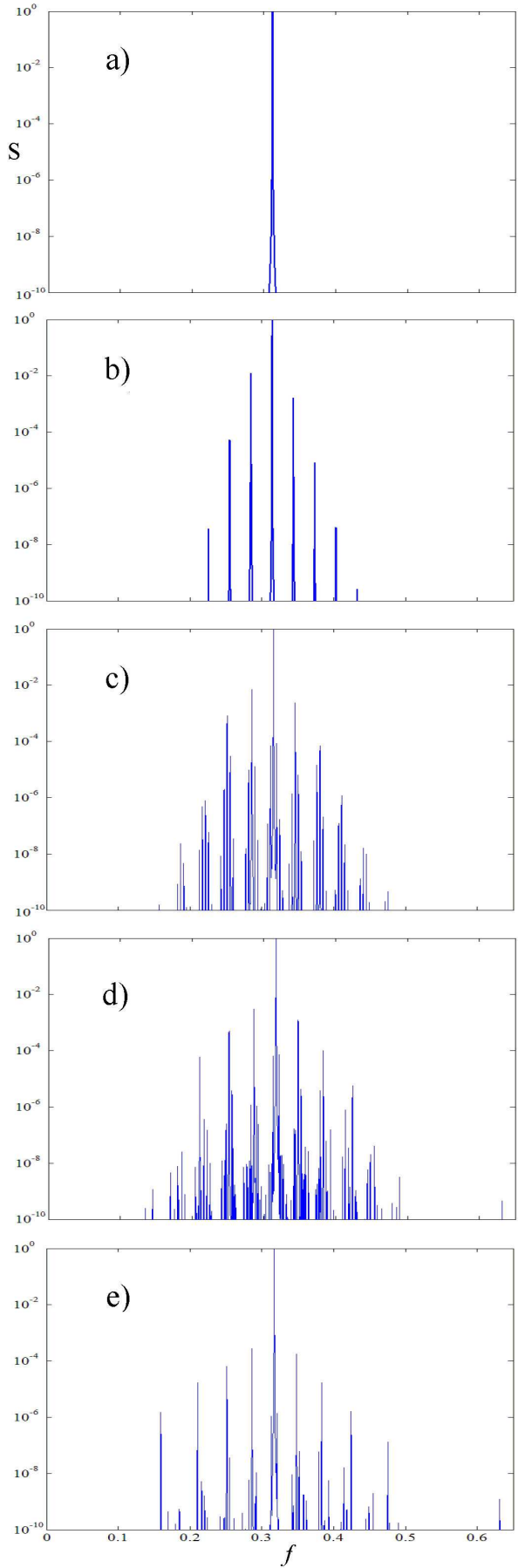}
\caption{The evolution of the Fourier spectrum fifth oscillator in (\ref{eq1_}) when the dissipation in the system,$\Delta =3$, a) $\mu =0.45$, b) $\mu =0.35$, c) $\mu =0.25$, d) $\mu =0.15$, e) $\mu =0.05$.}
\label{f4}
\end{figure}

Figures~\ref{f3} and~\ref{f4} illustrate the development of the Landau-Hopf scenario.

Fig.~\ref{f3} demonstrates evolution of the phase portraits of the oscillators with decrease of the dissipation parameter. Each diagram shows portraits for all five oscillators on their phase planes (the generalized coordinate versus the generalized velocity), and the orbits relating to the different oscillators are drawn in different colors. Fig.~\ref{f3}a corresponds to a large enough value of the dissipation parameter$\, \mu =0.45$. In this case we have a limit cycle embedded in the multi-dimensional state space of the system. It is seen clearly that the oscillations of the fifth oscillator are the most intense (it is so because the control parameter $\lambda$ of this oscillator is the largest). The intensity of the motion for the other oscillators decreases. In panel (b) corresponding to the three-frequency torus at$\, \, \mu =0.25$, the trajectories of all oscillators are perturbed quasi-periodically. However, in panel (c), at$\, \, \mu =0.05$, where the five-frequency torus occurs, the magnitude of the perturbations falls again: the small coupling weakly perturbs the orbits.

Figure~\ref{f4} shows evolution of the Fourier spectrum of the fifth oscillator in the course of development of the Landau-Hopf scenario in the system (\ref{eq1_}). One can observe the gradual enrichment of the spectrum with new spectral lines corresponding to the quasi-periodic regimes of the increasing dimension. Note meanwhile that in the panel (e) the degree of complexity of the spectrum visually rather decreases, although, strictly speaking, the dimension of the motion grows (the five-dimensional torus). The reason is the significant reduction in the level of coupling: at very low coupling the attractor is perturbed very weakly, and is close to the limit cycle of the autonomous oscillators (panel (c)). Therefore, the height of the corresponding spectral lines decreases, although in smaller scales the spectrum remains complex.

\begin{figure}
\includegraphics[width=0.45\textwidth,keepaspectratio]{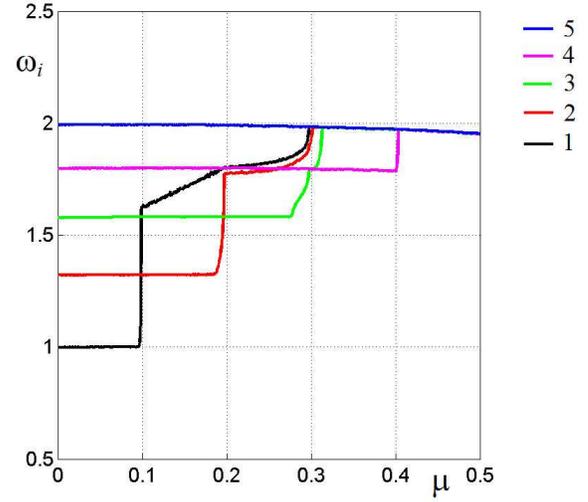}
\caption{The observed frequencies $\omega_i=<\dot{\varphi}_i>$ versus the coupling parameter $\mu$ for $\Delta=3$.}
\label{f5}
\end{figure}

Figure~\ref{f5} shows a plot for the observed frequencies   versus the coupling parameter   for the case $\Delta=3$. The frequencies are defined as the average rates of variation in time for the phases of the partial oscillators: $\omega_i=<\dot{\varphi}_i>$. They are evaluated numerically by counting a number of crossings of a certain surface in the phase space for a long time interval. The surfaces selected correspond to zero values for coordinate variables for the respective oscillators. From Figure~\ref{f5} one can see that in the range $0.4<\mu<0.5$, for the coupling level large enough, all the frequencies are the same, approximately equal to the frequency of the most excited fifth oscillator. With the decrease of the coupling strength, at $\mu\approx0.4$, a Neimark-Sacker bifurcation occurs; in Figure~\ref{f5} one can observe the appearance of one more frequency, which is close to the natural frequency of the fourth oscillator. Next, at $\mu\approx0.3$, one more new frequency branches out, which quickly approaches a value corresponding to the natural frequency of the third oscillator. A specific interesting behavior occurs for the remaining two oscillators. Their frequencies undergo a sharp change almost simultaneously with the third oscillator, but in the range $0.2<\mu<0.3$ they follow approximately the higher frequency of the fourth oscillator. In other words, the excited mode forces the weakly excited oscillators to adjust their frequencies to the former one. Then, the fourth frequency branches out, and, finally, at the parameter value $\mu\approx0.1$, the fifth frequency appears.\footnote{To some extent, anomalous is behavior of the first oscillator frequency in the parameter range $0.1<\mu<0.2$. The reason is that this oscillator is the least excited one. For it, the orbit is located entirely nearby the origin (Fig.~\ref{f3}a), and in this range of the coupling parameter the phase is poorly defined.} As we see, the picture of the "tree of synchronization" arises in the system manifesting the sequence of quasi-periodic Hopf-Neimark-Sacker bifurcations. This tree is of a rather specific sort, essentially distinct from that intrinsic to the phase oscillatory systems of traditional kind (see, for example, Figure 1.11a in ~\cite{6}, or Figure 1 in ~\cite{15}, etc.). In our case, one of the frequencies remains almost constant in the course of each bifurcation, while a new component branches out sharply. This is so because of the difference is the involved types of bifurcations: in Refs.~\cite{6,15} these are the saddle-node bifurcations of invariant tori, while in our case these are the secondary quasi-periodic Hopf-Neimark-Sacker bifurcations.

\subsection{Conclusion}

Thus, under certain conditions, e.g. in the case of non-identical parameters of the active modes and of detuning for the modes in ensemble of self-oscillating elements, one can observe a sequential cascade of soft quasi-periodic bifurcations involving tori of increasing dimension that can be regarded as occurrence of the Landau-Hopf scenario. The model outlined in this paper can be implemented, say, as an electronic device, which will serve as an example of a physical system that demonstrates this scenario under variation of the control parameter.

\section{Acknowledgement}

The work was supported by RFBR grant No 12-02-00541à and Grant Program of the Government of the Russian Federation for state support of scientific research conducted under the supervision of leading scientists at Russian institutions of higher professional education (Contract No11.G34.31.0039).

\textbf{\underbar{}}

\end{document}